\documentclass[10pt,conference]{IEEEtran} 
\usepackage{amsthm}

\usepackage{textcomp}

\usepackage{setspace}
\usepackage[]{changes}
\usepackage{latexsym}
\usepackage{balance}
\usepackage{graphicx,url}
\usepackage[caption=false]{subfig}

\usepackage[english]{babel}   
\usepackage[T1]{fontenc}

\usepackage{amsmath}
\usepackage{cases}

\usepackage{multirow}
\usepackage{booktabs}

\newcommand{\abstitle}[1]{\textbf{\textit{#1}}}

\usepackage{xcolor}
\usepackage{todonotes}

\newcommand{\repo}[1]{{\footnotesize{\sf{#1}}}}
\newcommand{\answer}[2]{{\em ``#1"}{ (#2)}}

\usepackage{xspace}
\newcommand{\tfdd}{TFDD\xspace}
\newcommand{\tfdds}{TFDDs\xspace}
\newcommand{\impr}{\mbox{\sl impr}}

\definecolor{gray}{RGB}{90,90,90}

\newtheoremstyle{def_style}% name of the style to be used
{\topsep}% measure of space to leave above the theorem. E.g.: 3pt
{\topsep}% measure of space to leave below the theorem. E.g.: 3pt
{}% name of font to use in the body of the theorem
{0pt}% measure of space to indent
{\itshape}% name of head font
{:}% punctuation between head and body
{ }% space after theorem head; " " = normal interword space
{\thmname{#1}\thmnote{(#3)}
}

\theoremstyle{def_style}

\newtheorem{example}{Example\ignorespaces}

\usepackage{mdframed}

\usepackage{enumitem}

\newmdtheoremenv[]{summary}{Summary\ignorespaces}

\usepackage{soul}
\usepackage{balance}
\usepackage{changepage}
\usepackage{framed}
\makeatletter
 {\par\unskip\endMakeFramed}
\makeatother

\definecolor{formalshade}{rgb}{0.93,0.93,0.93}

\definecolor{darkblue}{rgb}{0.2, 0.2, 0.2}

\newenvironment{formal}{%
  \def\FrameCommand{%
    \hspace{1pt}%
    {\color{darkblue}\vrule width 2pt}%
    {\color{formalshade}\vrule width 4pt}%
    \colorbox{formalshade}%
  }%
  \MakeFramed{\advance\hsize-\width\FrameRestore}%
  \noindent\hspace{-1pt}% disable indenting first paragraph
  \begin{adjustwidth}{}{7pt}%
  \vspace{2pt}\vspace{2pt}%
}
{%
  \vspace{3pt}\end{adjustwidth}\endMakeFramed%
}

\newcounter{resultcounter}

\newcounter{patterncounter}

\title{On the abandonment and survival of open source projects: An empirical investigation}

\author{\IEEEauthorblockN{Guilherme Avelino\IEEEauthorrefmark{1}, Eleni Constantinou\IEEEauthorrefmark{2}, Marco Tulio Valente\IEEEauthorrefmark{3}, Alexander Serebrenik\IEEEauthorrefmark{4}}
  \IEEEauthorblockA{%
    \IEEEauthorrefmark{1}Federal University of Piaui, Brazil, gaa@ufpi.edu.br
  }
  \IEEEauthorblockA{%
    \IEEEauthorrefmark{2} University of Mons, Belgium, eleni.constantinou@umons.ac.be
  }
  \IEEEauthorblockA{%
    \IEEEauthorrefmark{3} Federal University of Minas Gerais, Brazil, mtov@dcc.ufmg.br
  }
  \IEEEauthorblockA{%
    \IEEEauthorrefmark{4}Eindhoven University of Technology, The Netherlands, a.serebrenik@tue.nl
  }}

\begin{document} 

%%%%% UNCOMMENT FOR THE CR VERSION
 \IEEEoverridecommandlockouts
 \IEEEpubid{\makebox[\columnwidth]{978-1-7281-2968-6/19/\$31.00 \copyright 2019 IEEE \hfill} \hspace{\columnsep}\makebox[\columnwidth]{ }}

\maketitle
\begin{abstract}
\abstitle{Background}: Evolution of open source projects frequently depends on a small number of core developers. The loss of such core developers might be detrimental for projects and even threaten their entire continuation. However, it is possible that new core developers assume the project maintenance and allow the project to survive. 
\abstitle{Aims}: The objective of this paper is to provide empirical evidence on: 1) the frequency of project abandonment and survival, 2) the differences between abandoned and surviving projects, and 3) the motivation and difficulties faced when assuming an abandoned project. 
\abstitle{Method}: We adopt a mixed-methods approach to investigate project abandonment and survival. We carefully select 1,932 popular GitHub projects and recover the abandoned and surviving projects, and conduct a survey with developers that have been instrumental in the survival of the projects. 
\abstitle{Results}: We found that 315 projects (16\%) were abandoned and 128 of these projects (41\%) survived because of new core developers who assumed the project development. The survey indicates that (i) in most cases the new maintainers were aware of the project abandonment risks when they started to contribute; (ii) their own usage of the systems is the main motivation to contribute to such projects; (iii) human and social factors played a key role when making these contributions; and (iv) lack of time and the difficulty to obtain push access to the repositories are the main barriers faced by them. 
\abstitle{Conclusions}: Project abandonment is a reality even in large open source projects and our work enables a better understanding of such risks, as well as highlights ways in avoiding them.
\end{abstract}

%\alex{Fix xx\% throughout the manuscript}

\begin{IEEEkeywords}
Project abandonment, Truck factor, Bus factor, Open source development, Core developers
\end{IEEEkeywords}

%!TEX root = paper.tex
\section{Introduction}
\label{sec:intro}
Open source software (OSS) is crucial for society. Many proprietary software systems nowadays depend on open source frameworks and libraries, e.g., Instagram publicly acknowledges %and thanks 
the developers responsible for the open source libraries used in their site\footnote{\url{https://www.instagram.com/about/legal/libraries/}}.
Moreover, 72\% of GitHub survey participants report that they always seek out OSS options when looking for  tools\footnote{http://opensourcesurvey.org/2017/}. 
Importance of OSS also implies growing demands on sustainability of OSS projects. 
Sustainability of OSS projects is, however, a matter of concern since OSS projects are often managed by a small number of developers, without financial support~\cite{nadia2016roads}. 
For example, OpenSSL, a cryptography library used by two-thirds of all Web servers, was maintained by a single developer until 2014, when a major bug, nicknamed Heartbleed, affecting millions of sites was detected in its implementation~\cite{Durumeric2014}.

An easy way to communicate and understand the dependency of a software project on key developers is the notion of Truck Factor (TF), i.e., the minimal number of developers that the project depends on for its maintenance and evolution~\cite{Williams2003}. 
Stated otherwise, if the TF developers abandon the project (e.g., after being hit by a truck) the project maintenance will be heavily affected.
Recently, a number of researchers turned their eyes on the importance of studying the TF of software projects, specifically open source ones. 
Zazworka et al.~\cite{Zazworka2010} were the first to propose a heuristic to compute TFs by mining data from version repositories. Cosentino et al.~\cite{Cosentino2015} worked on a tool (and novel algorithm) for the same purpose, but targeting git-based repositories. Later, Avelino et al.~\cite{Avelino2016} proposed a heuristic to estimate TFs, based on a code authorship metric. 
However, the studies going beyond measuring TF towards more profound understanding of what happens when influential TF developers leave the project are still missing. 
We refer to such a situation as \textit{TF developers detachment} (\tfdd).

In this paper, we investigate \tfdd with the aim of identifying strategies that help projects to survive. 
We conduct a mixed-methods study following a sequential explanatory strategy~\cite{Easterbrook2008}. 
We start by collecting, curating, and analyzing a dataset of 1,932 popular GitHub projects. 
Using this dataset, we quantitatively address three research questions: \textbf{(RQ1)} How common are \tfdds in open source projects?, \textbf{(RQ2)} How often open source projects survive \tfdds? and \textbf{(RQ3)} What are the distinguishing characteristics of the surviving projects? These questions will shed light in the prevalence of \tfdds (\textbf{RQ1}), project survival (\textbf{RQ2}), and evolution of surviving and non-surviving projects (\textbf{RQ3}).

Next, we focus on the projects that survive \tfdds and survey 33 developers who assumed the maintenance of a studied project after it was abandoned by its original TF developers. 
Our qualitative investigation aims to answer three more research questions: \textbf{(RQ4)} Do new TF developers perceive risks of project discontinuation?, \textbf{(RQ5)} What motivates a developer to assume an open source project after a \tfdd situation? and \textbf{(RQ6)} What project characteristics most facilitate or hamper the work of recently arrived TF developers? 
We use this survey to provide qualitative answers about developers' awareness of \tfdd occurences (\textbf{RQ4}), their motivation to assume the responsibility for the project (\textbf{RQ5}), and enablers and barriers they have experienced while doing so (\textbf{RQ6}).

Our contributions are threefold. 
\emph{First}, we propose a methodology to identify \tfdds by mining software repositories and particularly to identify systems that survive (Section~\ref{sec:TF}).
\emph{Second}, we show that \tfdd is not just a theoretical concept.
%we provide a thorough characterization of 357 TF detachment occurences, detected during the evolution of 1,932 popular GitHub projects; we also show that only 41\% of the projects facing a TF detachment fully recover the maintenance work afterwards (Section~\ref{sec:res}).
\emph{Finally}, by surveying TF developers that assumed the maintenance of the surviving systems, we reveal their motivations and difficulties they faced when doing so.
%we reveal that 53\% were previously using these systems, i.e.,~they were motivated by their own need to fix bugs or to implement features; we also found that human and social characteristics had a key role to make the work of these contributors easier (Section~\ref{sec:survey}).

%!TEX root = paper.tex
\section{Truck Factor}
\label{sec:TF}
In this section, we first define concepts pertaining to TF. 
Then, we describe the approach used in the study to calculate TF, identify \tfdd and the systems that survived it.

%\subsection{Definitions}
\noindent The key definitions used throughout this paper are as follows:
\begin{itemize}[leftmargin=*]
\item {\em Truck factor (TF)} is the minimal number of developers of a project that have to be hit by a truck (or quit) before the project gets in serious trouble~\cite{Williams2003,Zazworka2010, Lavallee2015}. 
\item {\em TF developers} are the minimal set of developers $\{d_1, d_2, ..., d_n\}$ corresponding to TF. 
Typically, algorithms estimating TF also compute this set.
\item \textit{TF developers detachment} (\tfdd) occurs when all TF developers abandon the project. 
\item {\em Surviving system} is a system that survives a \tfdd, by attracting new TF developers who assume its maintenance.
\end{itemize}

\subsection{Truck Factor Calculation}
\label{sec:algorithm_tfcalculation}
To estimate truck factors we use the algorithm proposed by Avelino et al.~\cite{Avelino2016}. 
The selected TF algorithm initially calculates the degree of authorship (DOA). DOA~\cite{Fritz2010,Fritz2014} is a metric reflecting a developer's expertise on each file of the project relatively to the expertise of other developers on the same file. Expertise of a developer on a file is operationalized as the function of whether the developer has created the file, and the number of changes they did on the file compared to changes performed by other developers. 
Finally, TF estimation relies on the assumption that TF developers are the main authors, i.e., with the highest DOA, of at least 50\% of the system's files. 
We stress that there maybe more than one main author per file, as indicated in the TF algorithm description~\cite{Avelino2016}. 
The reasons for choosing this algorithm are fourfold:
(1) it has the best precision and recall in a recent study comparing three algorithms for estimating truck factors~\cite{Ferreira2017}; (2) it scales to large projects with hundreds of contributors; (3) it was validated by surveying the developers of 67 popular GitHub projects~\cite{Avelino2016}; (4) it has a public implementation on GitHub.\footnote{\url{https://github.com/aserg-ufmg/truck-factor}}

\subsection{Identifying Truck Factor Developers Detachments}
\label{sec:algorithm_tfEvent}
To search for \tfdds, we first estimate the TF of a system at a time $t$ and verify whether the TF developers abandoned the system before $t$. We say that a developer \textit{abandoned} a project if their last commit occurred at least one year before the most recent repository commit. 
Existing studies rely on different thresholds to classify developers inactivity or departure from a project, including three months~\cite{Constantinou2017}, six months~\cite{Lin2017, Foucault2015}, and one year~\cite{Izquierdo2009, Constantinou:2017:ISSE}. 
We experimentally test the sensitivity of five thresholds, in Section~\ref{sec:threshold}, and select the one-year threshold as it is the least sensitive to error.
%Since truck factor is viewed as a drastic event, we decide to be conservative and use a minimal one-year of inactivity period to define that a developer abandoned a project.
%To validate the threshold value selection, in Section~\ref{sec:threshold}, we test the sensitivity of five thresholds and experimentally show that the one-year value is the least sensitive to error.

\begin{figure}[!t]
    \centering
    \includegraphics[width=.9\linewidth]{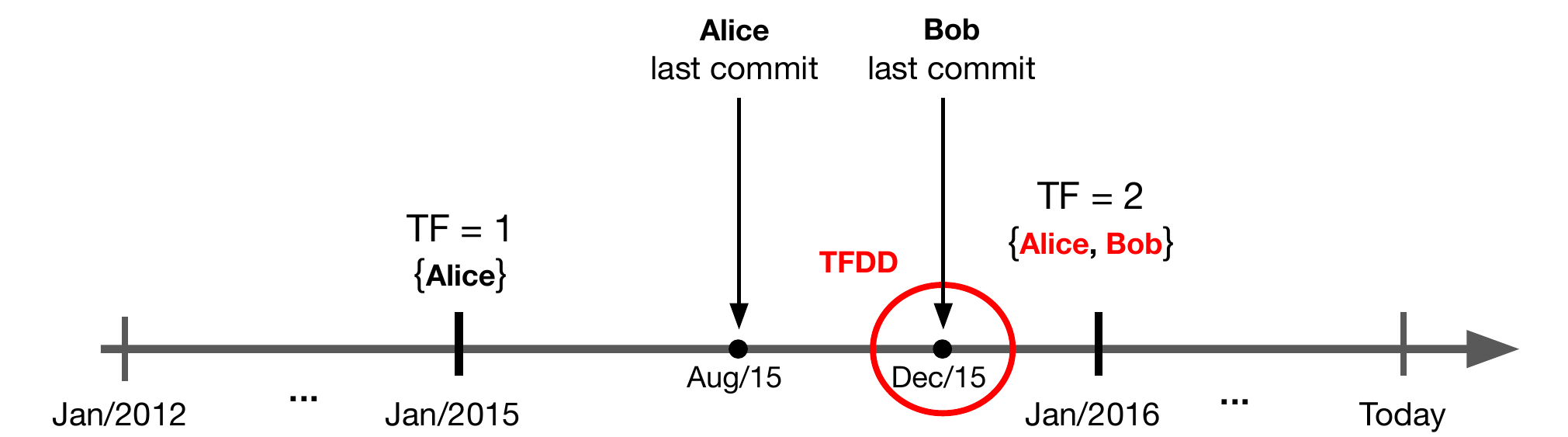}
    \caption{\tfdd on \textit{composer/satis}}
    \label{fig:dev_hist}
%    \eleni{Change the word event in the figure}
\end{figure}

%\alex{can we replace the real names of the developers with the fake ones to protect the privacy of the individuals involved?}
%\eleni{We need to: 1) state it clearly, 2) change the names with fictitious ones in the figures and 3) adapt the text. As soon as Guilherme decides what the fake names will be, we can adapt everything. For now, I will add the corrections in comments to remember that it is not yet finalized. We can still provide the repository link (I think so at least)}
%\alex{Why wouldn't we just call this Project1 and developers Alice, Bob and Charlotte?}
% \guilherme{Following Alexander's  suggestion I changed the names, as follow \\
%@Seldaek => Alice \\
%@JamesRezo => Bob \\
%@alcohol => Charlotte}

\begin{example}
For the sake of simplicity we do not reproduce the algorithm here, instead we illustrate how it is used in our context. 
Figure~\ref{fig:dev_hist} illustrates a fragment of the \repo{composer/satis}\footnote{\url{https://github.com/composer/satis}} development history\footnote{To preserve the privacy of the contributors involved, we replace their usernames with pseudonyms.}.
Suppose we first compute the system's TF in January 2015. 
At this point, the TF estimated by the algorithm equals one, since \textit{Alice} is the (unique) TF developer. 
As \textit{Alice} is active in January 2015 (she has a commit after this date), no \tfdd is observed.
When we compute TF in January 2016, TF increases to two, with \textit{Alice} and \textit{Bob} as the TF developers. 
Moreover, both developers abandoned the project before this date: \textit{Alice} in August 2015 (date of her last commit) and \textit{Bob} in December 2015. 
Therefore, the developers of \repo{composer/satis} detached from the project in December 2015. 
%Therefore, we say that  \repo{composer/satis} faced a TF event in December 2015. 
%We repeat this procedure in intervals of one year, aiming to check for TF events in multiple points of a project history. 
\end{example}
%We repeat this procedure in intervals of one year, aiming to check for TF events in multiple points of a project history. 
%\added{We investigate two different intervals: 3 and 6 months. With this changes the percentage of systems with TF events raise 3\% and 4\%, respectively for 6 and 3 months.  }
%\alex{How precise is our identification of the TF events? Not TF developers but \textbf{events}? This was one of the ICSME questions.}
%\eleni{What would be a feasible solution to answer this? I thought of a few things (e.g., examine the social continuity of the project after the TF event but it is not easy to do this within the day)}

\subsection{Identifying Surviving Systems}
\label{sec:algorithm_survivor}
By definition of TF, \tfdds are expected to have a major impact on the evolution of the software project. 
However, projects can survive such situations. 
In other words, an occurrence of \tfdd does not necessarily imply project termination, e.g., if new developers have taken charge of the project. 
%In contrast, when no significant development is performed after a TF event, the project is at risk since ``old'' TF developers have left it and ``new'' developers (if any) do not have enough expertise.
%Such projects are identified in our analysis since subsequent TF computations provide an identical set of TF developers as the ones of previous time periods.
%in these cases, further TF computations provide the same results of previous periods.

%\begin{newdefinition}
%Let $\mathit{TF}_1$, $\mathit{TF}_2$, ..., $\mathit{TF}_n$, be a sequence of TF developer sets of a system $S$, 
%where $\mathit{TF}_t$ is the TF developer set computed at time $t$. 
%We say that $S$ \emph{survived} $\mathit{TF}_{t_1}$ if there exists $t_2$, $1\leq t_1 < t_2\leq n$, such that
%\begin{itemize}
%    \item[(a)] all $d\in \mathit{TF}_{t_1}$ abandoned the project before $t_1$, and
%    \item[(b)] $\mathit{TF}_{t_2}\setminus \mathit{TF}_{t_1} \ne \emptyset$, i.e.,~at the time interval $[t_1, t_2]$ at least one new developer performed 
%\deleted{important} 
%\added{a significant number of}
%a significant number of 
%contributions to the point of entering in $\mathit{TF}_{t_2}$.
%\end{itemize}
%\end{newdefinition}

We assume a project can be in two states: {\em Active}, when at least one TF developer is active; and {\em Inactive}, when all TF developers have abandoned the project. 
When a \tfdd occurs, the system is moved from {\em Active} to {\em Inactive}; reversely, the attraction of at least one new TF developer moves the project back to {\em Active}. 
Our central object of study are systems with a transition from {\em Inactive} to {\em Active} w.r.t. the last occurrence of \tfdd; such systems are considered as having {\em survived} since they became active after their \emph{last} \tfdd.
% \eleni{@Guilherme: Is the last sentence true? I thought that the following is true:\\
% "In this paper, our central object of study are systems with a transition from from {\em Inactive} to {\em Active} w.r.t their last TF event; such systems are considered as having {\em survived} since they became active after their last TF event"}
% \guilherme{@Eleni, you are right. I changed the text. Please check.}

\begin{figure}[!t]
    \centering
    \includegraphics[width=.9\linewidth]{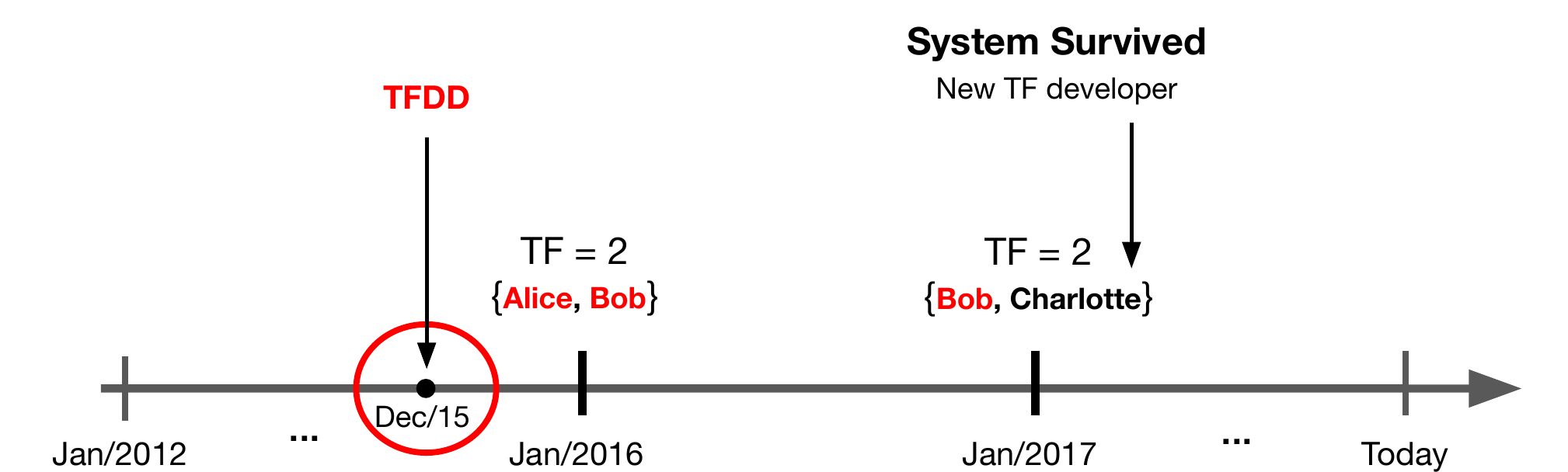}
    \caption{Surviving \tfdd on \textit{composer/satis}}
    \label{fig:survivor}
%     \eleni{Change the word event in the figure}
\end{figure}

\begin{example}
As illustrated in Figure~\ref{fig:survivor}, a \tfdd occurs in our running example (\repo{composer/satis}) on December 2015, when both TF developers abandoned the project. 
Therefore, in this date, the project moved to an {\em Inactive} state. 
However, in January 2017, the recomputation of the TF developers resulted in a new developer in this set, {\em Charlotte}.\footnote{We compute TFs every year, starting from the repository creation date.} The attraction of this developer---someone with important contributions to the point of reaching a TF status---moves the project back to the {\em Active} state. Thus, we say \repo{composer/satis} survived the \tfdd.
\end{example}

%!TEX root = paper.tex
\section{Study Design}
\label{sec:studyDesign}

%\eleni{Papers should contain an explicit description of the empirical strategy used or investigated.}

We adopt a mixed-methods approach and combine a large scale analysis of version control repository data with a survey. 
Mixed-methods are appropriate for the pragmatic stance common in software engineering research~\cite{Easterbrook2008}.
%, and were often applied in the past~\cite{Easterbrook2008}. 
 
\subsection{Dataset \& Preprocessing}
\label{sec:dataset}
To perform the quantitative part of the study, we build a dataset with GitHub projects. 
Initially, we focus on six programming languages with the largest number of GitHub repositories: JavaScript, Python, Ruby, C/C++, Java, and PHP. 
We select the top-500 most starred repositories (excluding forks to avoid including the same project multiple times) for each of those languages at the moment of analysis. 
We focus on popular projects to ensure the quality of the data, so that the collected projects are relevant to the OSS community, and to avoid including personal projects in our dataset~\cite{Kalliamvakou2015,MunaiahKCN17}.

To safeguard the quality of the dataset we filter the resulting collection of 3,000 GitHub repositories. 
We explicitly address well-known ``perils of mining GitHub''~\cite{Kalliamvakou2015}.
We exclude 
(a) projects that did not use GitHub exclusively during their entire history and lost part of their development history when migrated to GitHub, 
(b) projects that do not have sufficient historical data for the TF computation, and 
(c) projects that are not software units or are explicitly labeled as unmaintained. 
To identify projects with evidence of loss of part of their development history we filter out repositories where more than 50\% of the files are added in less than 20 commits in the beginning of their development. 
By applying this filter, we exclude 677 projects. 
As our approach to identify \tfdd requires at least two years of historical data, we filter out 338 projects with less than two years of development activity. 
To apply the last filter we manually inspect the project descriptions and exclude 53 projects.  Among others, we found repositories containing books, awesome-lists (i.e.,~sets of suggested books, links, etc.), and technology code samples. 
The resulting dataset is composed of 1,932 ($= 3,000 - 677 - 338 - 53$) projects. 

\begin{table}[!t]
  \centering
  \small
  \caption{Number of projects by language.}
  \label{tab:rep_by_languages}
   \scalebox{0.85}{
  \begin{tabular}{lr || lr}
  \toprule
  \multicolumn{1}{l}{\bf Language} & {\bf Projects}  &  \multicolumn{1}{| l}{\bf Language} & {\bf Projects}\\
  \midrule 
Ruby      & 398 (21\%) &  PHP         &  334 (17\%)  \\
JavaScript & 342 (17\%) &   Python & 297 (15\%)\\
C/C++         & 335 (17\%) & 	Java & 226 (12\%) \\
  \bottomrule
  \end{tabular}
  }
\end{table}

%\begin{figure}[!t]
%    \centering
%    \includegraphics[width=\linewidth]{figures/dataset_by_language.pdf}
%    \caption{Number of projects by language.}
%    \label{fig:rep_by_languages}
%\end{figure}

\begin{figure}[!t]
    \centering
    \includegraphics[width=.95\linewidth]{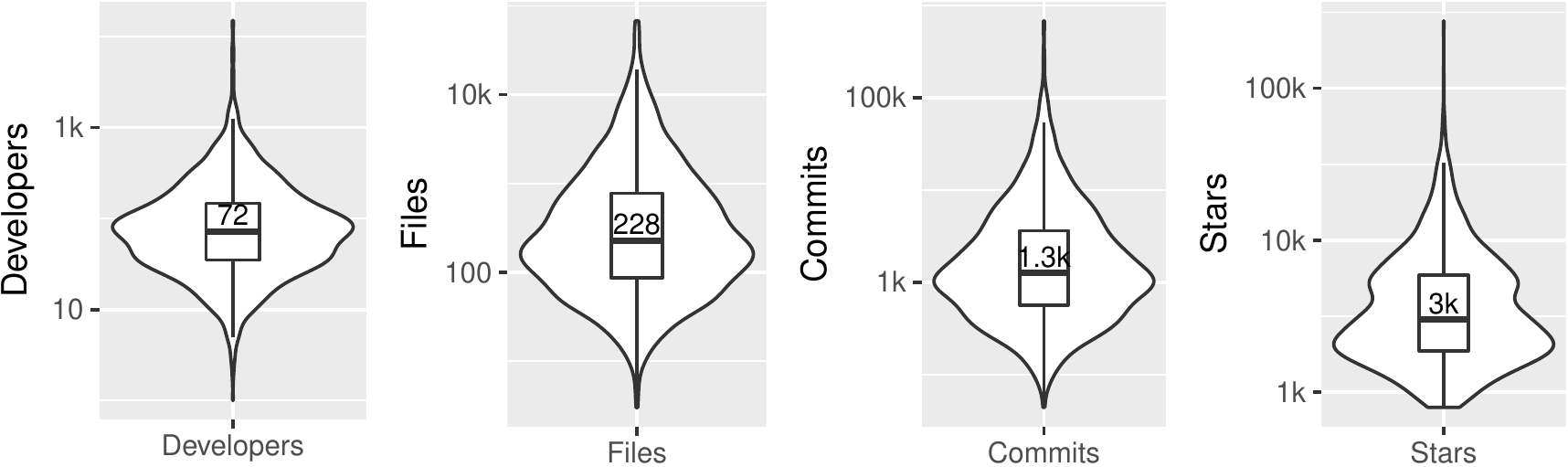}
    \caption{Distribution of the number of developers, commits, files, and stars.}
    \label{fig:dataset}
\end{figure}

As shown in Table~\ref{tab:rep_by_languages}, %Figure~\ref{fig:rep_by_languages}, 
most projects are implemented in Ruby (398 projects, 21\%). On the other side, Java is the language with fewest projects (226 projects, 12\%). 
Figure~\ref{fig:dataset} shows violin plots with the distribution of the number of developers, source code files, commits and stars per project (please note the logarithmic scale). 
The median values are indicated inside the violin plots. 
We conclude that the dataset constructed typically includes large systems, both in size and in number of developers, and that the systems also are popular (number of stars) and have a large number of commits. 

\subsection{Aliases Resolution} 
\label{sec:aliases}
\begin{figure}[!t]
   \centering
   \includegraphics[width=.85\linewidth]{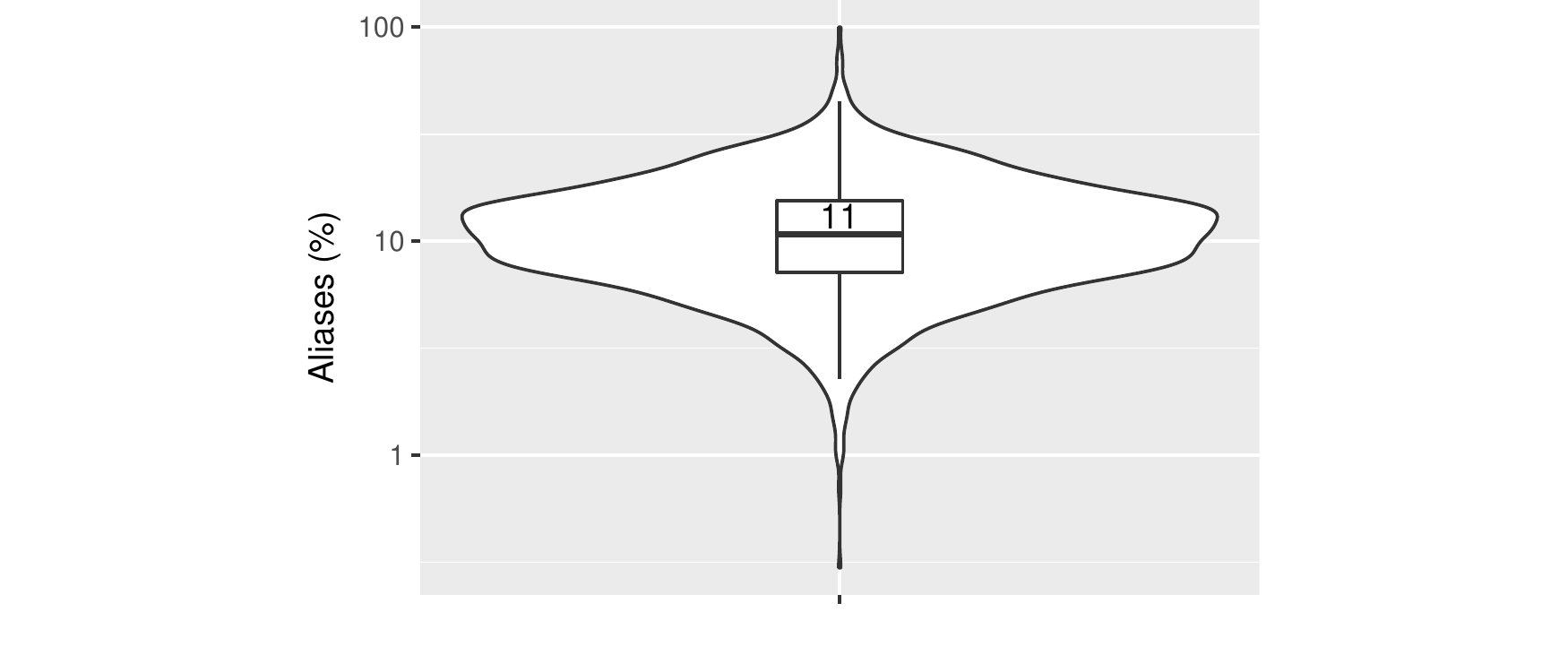}
   \caption{Percentage of aliases in each project}
   \label{fig:aliases}
\end{figure}

The correctness of TF computations highly depends on the set of distinct developers. 
However, developers do not necessarily use only one alias (name or e-mail address) when contributing to a project~\cite{Kouters2012,Goeminne2013,%
WieseSSTG16}. 
Therefore, it is important to detect and resolve aliases among the developers of the 1,932 projects in our dataset. 
Rather than using heuristics advocated in previous works to detect aliases~\cite{Kouters2012,Goeminne2013,WieseSSTG16}, we use a feature of the GitHub API that maps an e-mail address in the commit header to a GitHub user. 
Using this feature, we mapped each developer of each system to their GitHub account; $d_1$ and $d_2$ are considered aliases when they are mapped to the same GitHub account. 
As a downside, this approach does not handle the cases where developers have multiple GitHub accounts. 
Figure~\ref{fig:aliases} shows a violin plot with the percentage of aliases in each project. The median percentage of aliases in a project in our dataset is 11\%.

\subsection{Abandoner Threshold Sensitivity Analysis}
\label{sec:threshold}
The selected threshold value to identify developers \textit{abandoning} projects can impact our results. We therefore test the sensitivity to error of five different threshold values, i.e., 3 months, 0.5 year, 1 year, 1.5 year and 2 years, to select the most appropriate threshold.  
We first gather each TF developer's commit activity and then measure the elapsed time delta between consecutive commits. 
For each TF developer with $N$ commits, we compute $N-1$ inter-commit time deltas. 
Since each time delta represents the time elapsed between commits, developers should never be classified as abandoners by a threshold. 
In other words, appropriate thresholds should optimally have zero error, meaning that they will never erroneously classify a developer as an abandoner, as by definition she has at least one subsequent commit.

To assess the error sensitivity of a list of thresholds $\mathit{TS}$ where $\mathit{TS}$ = $\langle T_1, T_2, ..., T_N \rangle$ such that $T_i < T_{i+1}$ for $i \in 1...N-1$, we used the precision and improvement metrics, as well as their harmonic mean. Precision $P(T_i)$ of a threshold $T_i$ is defined as the percentage of developers that $T_i$ has zero error, i.e., $T_i$ never classifies them as abandoners. Improvement $\impr(T_i,T_{i-1})$ of $T_i$ over the smaller threshold $T_{i-1}$ is defined as the number of developers that $T_i$ has zero error, while $T_{i-1}$ erroneously classifies as abandoners over the total number of developers that $T_{i-1}$ erroneously classifies as abandoners. In practice,   $\impr(T_i,T_{i-1})$ measures how many errors of $T_{i-1}$ were corrected by $T_i$. The harmonic mean between precision and improvement is defined as $\frac{2*P*\impr}{P + \impr}$.

%\vspace{-0.3cm}
%\begin{align*}
%harmonic\_mean(P,impr) = \frac{2}{\frac{1}{P} + \frac{1}{impr}}
%\end{align*}
%\guilherme{@Eleni, I know these two equations are correct, but I believe this second option is more common and easy to understand. What do you think?}
%\begin{align*}
%\mbox{\sl harmonic}\_\mbox{\sl mean}(P,\impr) = \frac{2*P*\impr}{P + \impr}
%\end{align*}

\begin{table}[!t]
 \caption{Threshold sensitivity}
 \centering \small
 \scalebox{0.85}{
 \begin{tabular}{ c  r r r}
 \toprule
 \boldmath{$T_i$} & \boldmath{$P(T_i)$} & \boldmath{$\impr(T_i,T_{i-1})$} & \boldmath{$\mbox{\sl harmonic}\_\mbox{\sl mean}$}\\ 
 \midrule
 3 months & 0.38  & - & - \\  
 6 months & 0.59  & 0.35 & 0.44 \\  
 1 year & 0.82  & 0.55 & \textbf{0.66} \\  
 1.5 year & 0.91  & 0.50 & 0.64 \\  
 2 years & 0.95  & 0.46 & 0.62\\  
 \bottomrule
 \end{tabular}
 }
 \label{tab:thres_sens}
\end{table}

Table~\ref{tab:thres_sens} presents the sensitivity analysis results for the five threshold values considered. 
The precision results indicate that a certain amount of error is introduced regardless of the threshold, e.g., even a 2-year threshold produced an error of 5\%. 
On the contrary, the largest improvement is achieved by the 1-year threshold (55\%) over the 6-months threshold, indicating that more than half of the errors made by the 6-month threshold were fixed by the 1-year threshold.
Overall, the 1-year threshold achieves the highest harmonic mean value (66\%) compared to the other thresholds. We therefore use the 1-year threshold in our experiments to determine if a developer has abandoned the project after their last commit.

\section{Searching for \tfdds and Surviving Projects}
\label{sec:res}

\begin{figure}[!t]
   \centering
  \includegraphics[width=.89\linewidth]{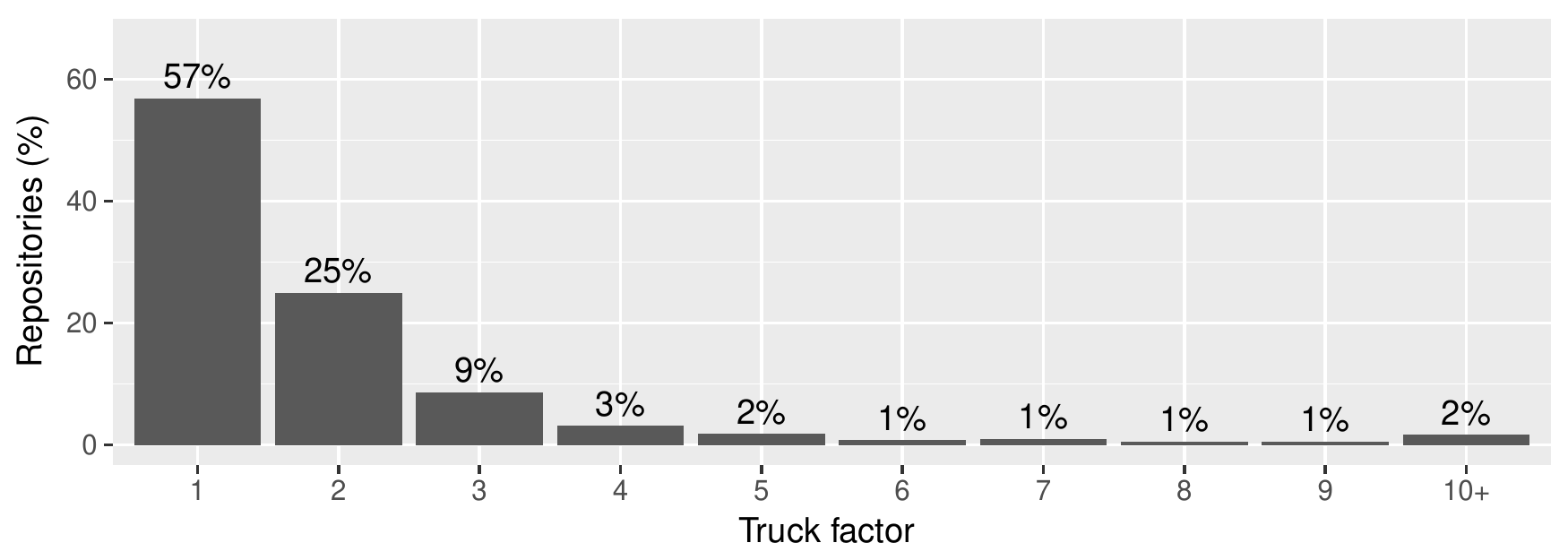}
   \caption{TF of the 1,932 projects in our dataset}
   \label{fig:dataset_TF}
\end{figure}

Prior to analyzing \tfdds, we estimate the TFs for 1,932 projects in our dataset using the algorithm of Avelino et al.~\cite{Avelino2016}. 
We clone the project repositories and hereby provide statistics based on the most recent snapshot of the considered repositories; the TF analysis is performed yearly since the first commit of each project to answer the first three research questions. 
%We clone the project repositories and analyze the most recent version at the moment of analysis. 
%\eleni{Reviewer:\\
%The results shown in Figure 8 ("TF of the 1,932 projects in our dataset") seem to be computed on the most recent releases of the selected projects. How about the rest of the results? Are the TFs computed on the basis of releases? If not, which time points are chosen for the computation?
%The validity of the TF, TF developers and TF events determines the validity of the results, it is extremely important to clarify the calculation.
%}
Figure~\ref{fig:dataset_TF} presents a histogram with the TF results. 
As we can observe, most projects have a low TF: 
e.g., for 57\% projects TF equals 1, while less than 6\% have a TF higher than 5. 
The highest TF is 26, computed for \repo{edx/edx-platform}, which is the software platform that supports edX massive open online courses. 
Our findings concur with the earlier results of
Avelino et al.~\cite{Avelino2016} that reported that 65\% of the evaluated systems have TF $\leq$ 2, based on a sample of 133 popular GitHub projects.

\begin{figure}[!t]
    \centering
    \includegraphics[width=.89\linewidth]{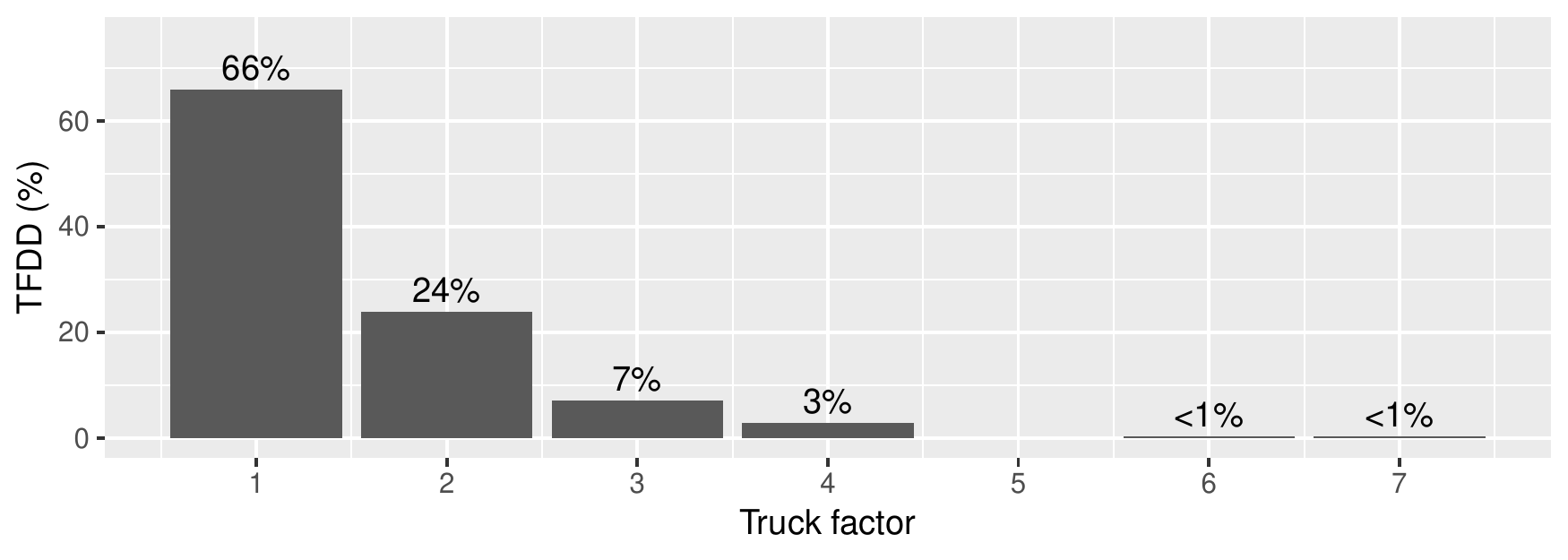}
    \caption{Projects facing \tfdds}
    \label{fig:tf-events}
%    \eleni{Fix the y axis label}
\end{figure}

\begin{formal}
Most open source projects have low TFs. In a sample of 1,932 projects, 57\% have TF = 1 and 25\% have TF = 2. The highest TF in our sample is 26 developers.
\end{formal}

In the remainder of this section, we describe a quantitative exploration of the collected data, aiming to answer \emph{(RQ1)}--\emph{(RQ3)}. 
We start by assessing whether \tfdds indeed happen in open source development \emph{(RQ1)}. 
Assuming that \tfdds indeed occur, \emph{RQ2} takes a step further and investigates how often projects overcome such situations. 
Finally, assuming we find projects that survived their \tfdds, we compare them with other projects that did not have the same fate \emph{(RQ3)}. The goal is to identify characteristics that might help projects to overcome the loss of  TF developers.
%\noindent {\em (RQ1) How common are truck factor events in open source projects?} To start our investigation, we assess whether TF events indeed happen in open source development.

%\noindent {\em (RQ2) How often open source projects survive a truck factor event?} Assuming the previous question reveals that TF events indeed occur, this second question takes a step further and investigates how often projects overcome such events.

%\noindent {\em (RQ3) How surviving projects differ from non-surviving ones?} Finally, assuming we find projects that survived their TF events, we compare them with other projects that did not have the same fate. The goal is to identify characteristics that might help projects to overcome the loss of  TF developers.

\subsection*{RQ1) How common are \tfdds in GitHub projects?}\label{sec:tfevents}
We identify \tfdds in 315 projects, 16\% of our dataset. 
Most of the projects faced only one \tfdd situation (88\%). However, some projects faced two (11\%) or even three ($<$ 0.1\%) \tfdds. 
Figure~\ref{fig:tf-events} shows the percentage of \tfdds grouped by TF. 
As expected, most \tfdds are observed in systems with a small TF, e.g., 66\% of \tfdds happens in projects with a TF equal to one. 
This means that most projects that are in a \tfdd situation are maintained by one core developer; it remains to be seen if most projects are in such a situation only once because they become obsolete or because they survive it and never face one again. 
We further investigate project survival after \tfdds in Section~\ref{sec:survival_rate}.

\begin{figure}[!t]
    \centering
    \includegraphics[width=.99\linewidth,height=2.4cm]{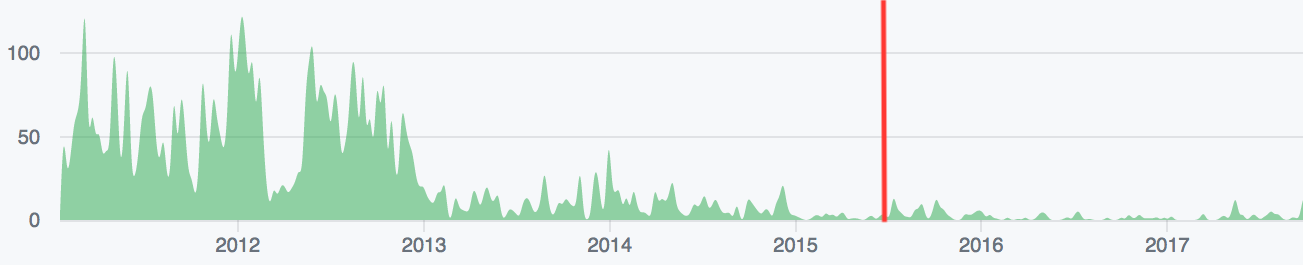}
    \caption{Contributions to PointCloudLibrary/pcl over time (screenshot from GitHub). A \tfdd occurred at June, 2015 (vertical red line).%, according to the proposed methodology.
    }
    \label{fig:pcl_example}
\end{figure}

In contrast, projects found in a \tfdd situation only twice have a TF higher than four: \repo{etsy/logster} ($\mathit{TF} = 7$) and \repo{PointCloudLibrary/pcl} ($\mathit{TF} = 6$). \repo{etsy/logster} is a small project, with 13 files and 117 commits when the \tfdd was observed.  
By contrast, \repo{PointCloudLibrary/pcl} is a large project, with 9,568 commits and 2,204 files at \tfdd time. 
All TF developers started contributing to this project in the first year of its development (2011), but abandoned the project before 2015. 
To show the impact of their departure, Figure~\ref{fig:pcl_example} shows a screenshot with the contributions to \repo{PointCloudLibrary/pcl}, as available on its GitHub page\footnote{https://github.com/PointCloudLibrary/pcl/graphs/contributors}.
Most contributions happened before June, 2015, when the project faced a \tfdd (vertical red line, in the figure). 
This was the date of the last commit of one of the TF developers. 
The commits of the other five TF developers all happened before May, 2014.
Although \repo{PointCloudLibrary/pcl} has had financial support from a non-profit organization,\footnote{http://www.openperception.org} as indicated in the project's README page, the site and social network accounts of this organization do not receive updates since 2014, which is close to the \tfdd date.

\begin{formal}
Truck Factor developers detachment is not merely a theoretical concept:  
16\% of the projects faced at least one \tfdd; 66\% of these \tfdds happened in systems with TF=1,  which are 55\% of the projects.
\end{formal}

\begin{figure}[!t]
    \centering
    \includegraphics[width=.89\linewidth]{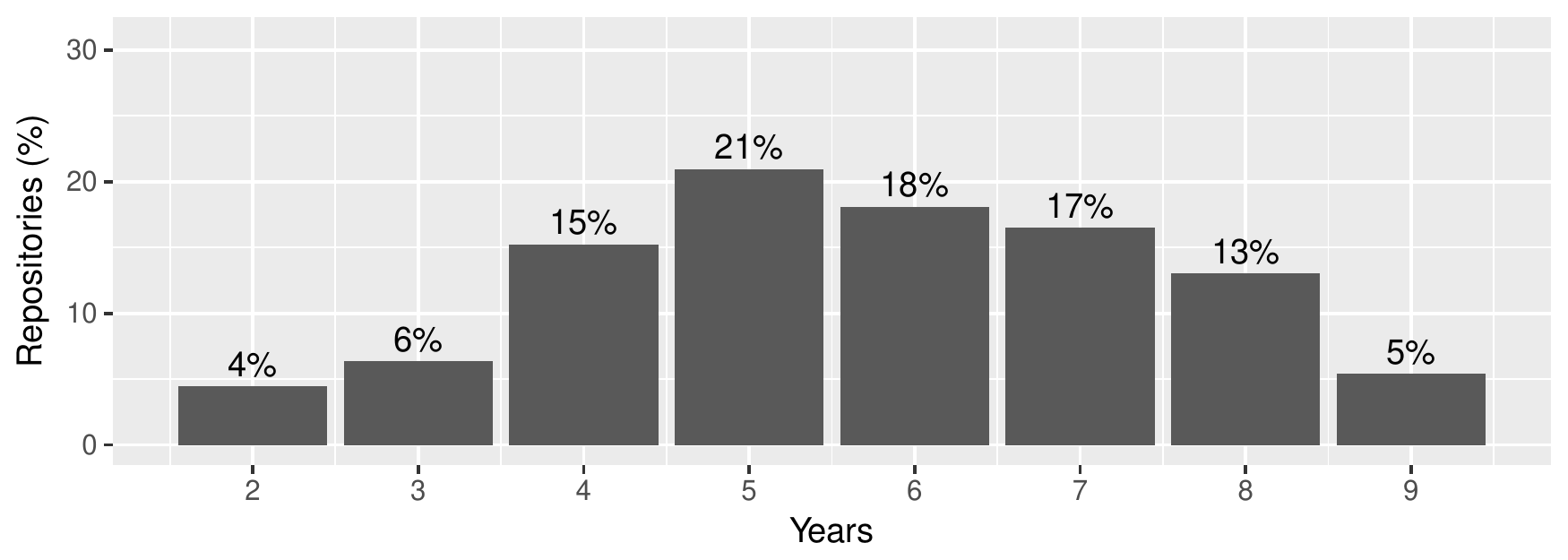}
    \caption{Age of the repositories with \tfdds}
    \label{fig:lifetime-tf_systems}
\end{figure}

\begin{figure}[!t]
    \centering
    \includegraphics[width=.89\linewidth]{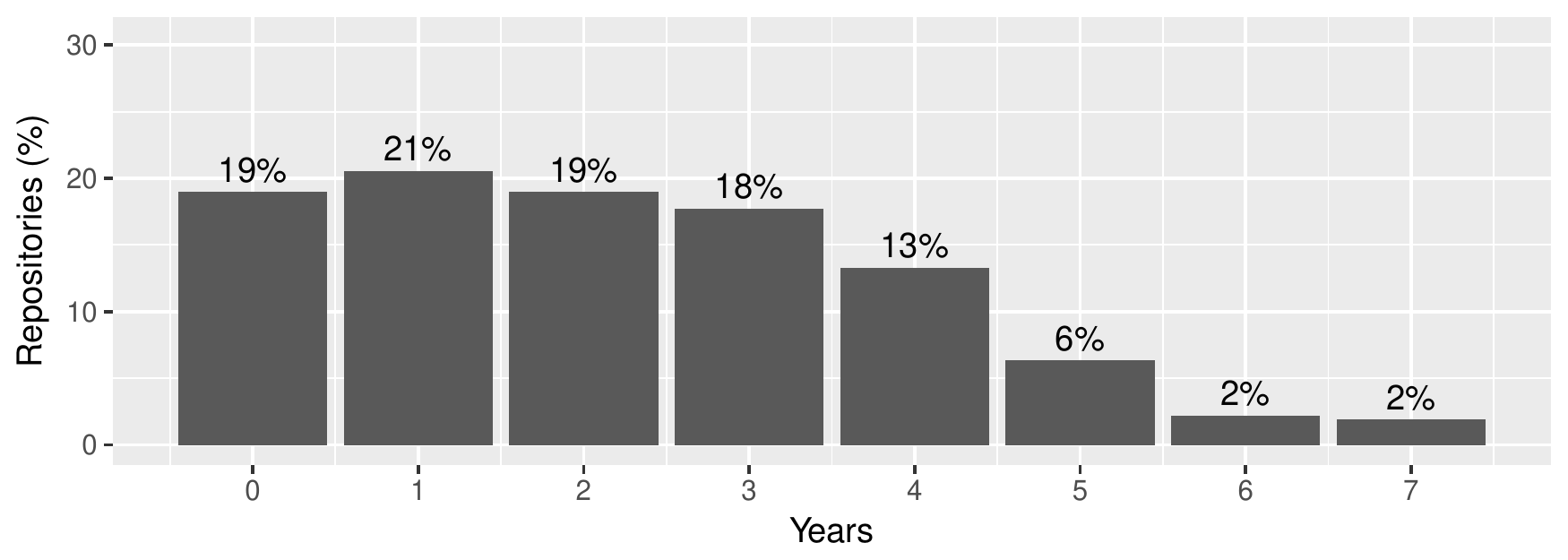}
    \caption{When do \tfdds happen (counting from the repositories creation)}
    \label{fig:lifetime-tf_event}
\end{figure}

Figure~\ref{fig:lifetime-tf_systems} shows the age of the repositories with \tfdds, considering their creation date on GitHub. 
As we can see, most projects (71\%) have between 4 and 7 years of development. 
Figure~\ref{fig:lifetime-tf_event} shows when these \tfdds happen, in terms of number of development years and counting only the first \tfdd, for projects with multiple \tfdds. 
As we can observe, there is a concentration of \tfdds in the first years of development; 59\% took place in the
first two years of development. 
In fact, in some cases the TF developers abandoned the projects some time after the repository creation, e.g., in 23 projects the
TF developers abandoned the projects in the first six months.

\begin{formal}
59\% of the \tfdds happened in the first two years of development; but 71\% of the projects with \tfdds have now between 4 and 7 years of development.
\end{formal}

%\begin{summary}
%    16\% of the studied projects faced at least one TF event. Most events (66\%) occurred in projects with TF = 1 and in the first two years of development (59\%).
%\end{summary}

%\begin{figure}[!t]
%   \centering
%   \includegraphics[width=.89\linewidth]{figures/survive_computations.pdf}
%   \caption{When do projects survive a \tfdd}
%   \label{fig:survive_computations}
%%   \eleni{fix the x axis label}
%\end{figure}

\subsection*{RQ2) How often open source projects survive a \tfdd?}\label{sec:survival_rate}
A project survives if it survives the last observed \tfdd.
In total, 128 projects (out of 315 projects) overcome their \tfdds, which represents a survival rate of 41\%. 
In most cases (86\%) we detected that only one new TF developer was attracted to the project and was responsible for its survival. However, there are cases where two (12\%) or even three (2\%) new TF developers were attracted to the projects. 
%\eleni{As a reviewer I would like some more information: new TF developers were not core developers before taking over the project. Were they never active before OR were they peripheral developers before becoming new TF devs?}
%\guilherme{The following discussion about the percentage of newcomers address this question.}
%Additionally, in 64\% of these cases the attraction occurred in the first year after the \tfdd, as presented in Figure~\ref{fig:survive_computations}. 
Additionally, in 64\% of these cases the attraction occurred in the first year after the \tfdd, while 23\% occur in the second year, 10\% in the third year and 2\% in the fourth year.
As expected, it becomes more difficult to attract new TF developers to assume project maintenance throughout the years. 
%In the meantime, projects either remain inactive or their evolution only relies on minor contributors.% as there are no core developers present to manage and lead the project. 

%there are no important developers present to manage the development tasks and therefore, peripheral developers can be working without any coordination. 
%\eleni{Needs to be improved}
%\deleted{As expected, it is more difficult to recover project maintenance after years of inactivity.}

\begin{formal}
It is possible to recover from \tfdds:
41\% of the projects survived their last observed \tfdd, usually by attracting a single new TF developer (86\%).
\end{formal}

A developer is called a \textit{newcomer} if their first commit occurs after the last observed \tfdd. Otherwise, they are an \textit{old-contributor}. In most surviving projects (52\%), the new TF developers are all \textit{old-contributors}. However, a significant part of the projects survived with the help of \textit{newcomers} (41\%) or by attracting both newcomers and old contributors (7\%).

\begin{formal}
Newcomers are crucial to recover from \tfdds. They contributed to recovery of 48\% of the surviving projects.%
%\eleni{I would mention as a finding that we cannot distinguish between old project contributors and newcomers since both have equally contributed to projects overcoming TF events}
\end{formal}

%%%%%%%\TODO: to better investigate alexander suggestion %%%%%
%\alex{* RQ2. Again baseline issue here. There are studies of the motivation of open source developers in general: are our new contributors somehow different from OS developers in general in this respect? This is an important point to check.}

\subsection*{RQ3) How surviving projects differ from non-surviving ones?}\label{sec:survival_non-surviving}

\begin{figure}[!t]
\centering
\includegraphics[width=.89\linewidth]{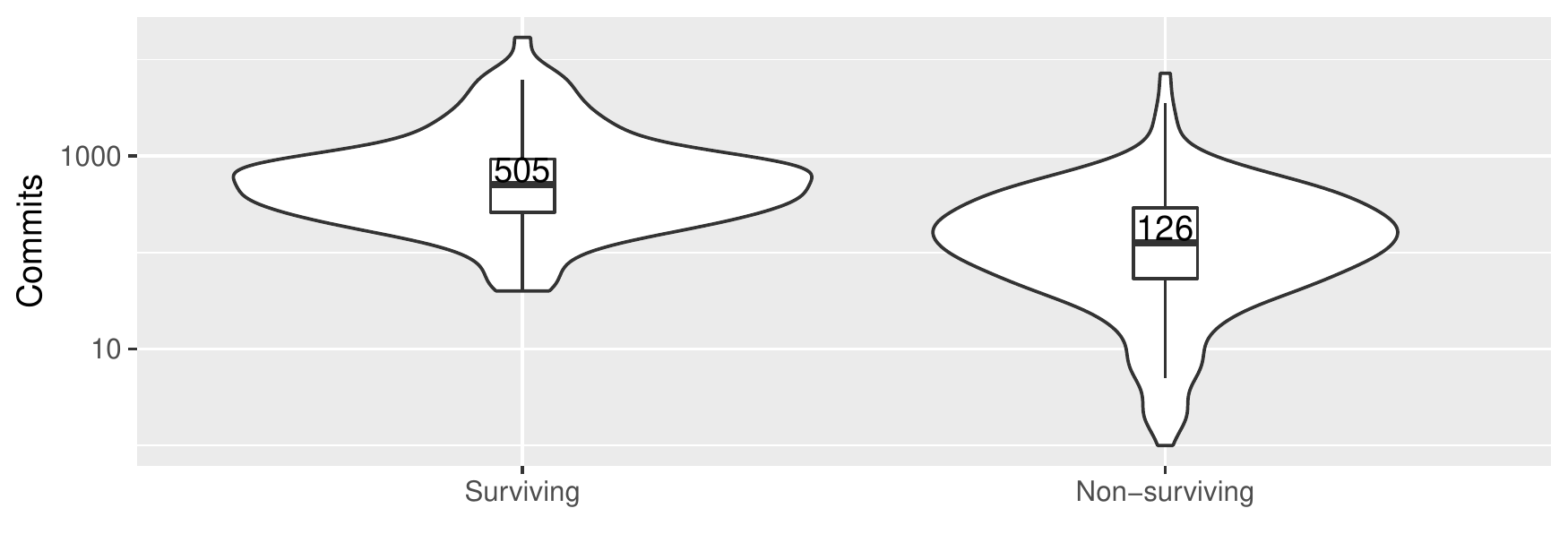}
\caption{Number of commits after the last observed \tfdds}
\label{fig:abs_commits}
\end{figure}

\begin{figure}[!t]
\centering
\includegraphics[width=.87\linewidth]{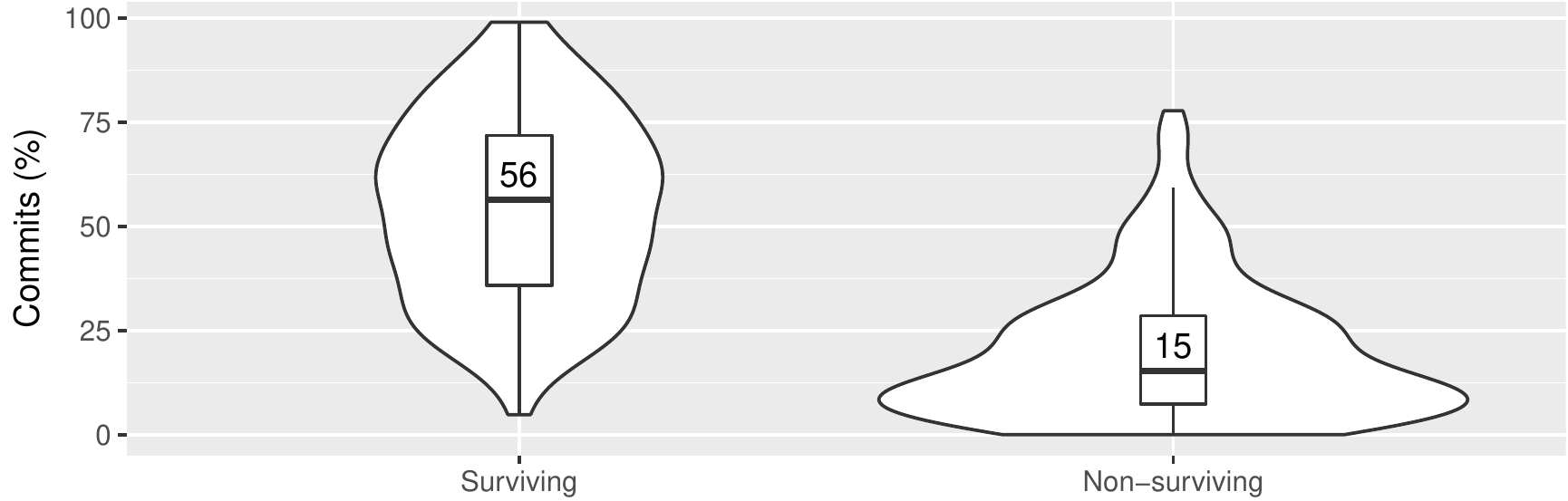}
\caption{Percentage of commits after the last observed \tfdds}
\label{fig:perc_commits}
\end{figure}

Figures~\ref{fig:abs_commits} and \ref{fig:perc_commits} show respectively the distribution of the absolute number and the percentage of commits after the last detected \tfdd in each surviving project (128 projects) and also in the non-surviving ones (187 projects). 
%For example, suppose a project with 100 commits, 95 until the date of its last TF event and 5 commits after this date; in this case, the percentage of commits after the studied TF event is 5\%. 
Before discussing these figures, we stress that \tfdd should have a major impact on project maintenance and evolution, but this does not necessarily mean that the project maintenance has ceased afterwards. 
Therefore, it is possible to observe commits after \tfdds even in non-surviving systems. However, these commits are 
\emph{performed by minor contributors} and do not affect the TF developer set. 
This means that the projects continue to be at risk even in the presence of commits after a \tfdd. 

%\eleni{Please check the paragraph below}
%We expect surviving projects to have more intense commit activity after the last \tfdd compared to the non-surviving ones as their maintenance is resumed by new TF developers. 
The violin plots in Figures~\ref{fig:abs_commits} and \ref{fig:perc_commits} show a clear difference between surviving and non-surviving systems. 
The surviving systems have 505 commits (56\%) after the last detected \tfdd, whereas the non-surviving ones have only 126 commits (15\%), considering the median values. 
The third quartile measures are 949 commits (72\%) and 289 commits (29\%) for surviving and non-surviving projects respectively. 
%These differences are confirmed using the one-sided version of the Mann-Whitney test (p $\leq$ 5\%). 
These differences are confirmed using the one-sided version of the Mann-Whitney test ($p=5.02\times10^{-22}$ and $p=2.04\times10^{-32}$ for the number and percentage of commits after the last \tfdd respectively). 
The effect size, measured by  Cliff's delta~\cite{Grissom2005} and using the intervals of Romano et al.~\cite{Romano:2006}, is \textit{large} in both cases: $d=0.64$ for the number of commits, and $d=0.79$ for the percentage of commits after the last \tfdds.
%\eleni{
%Commits before: p$<$0.01, d=0.25, small effect}

\begin{figure}[!t]
    \centering
    \includegraphics[width=.97\linewidth]{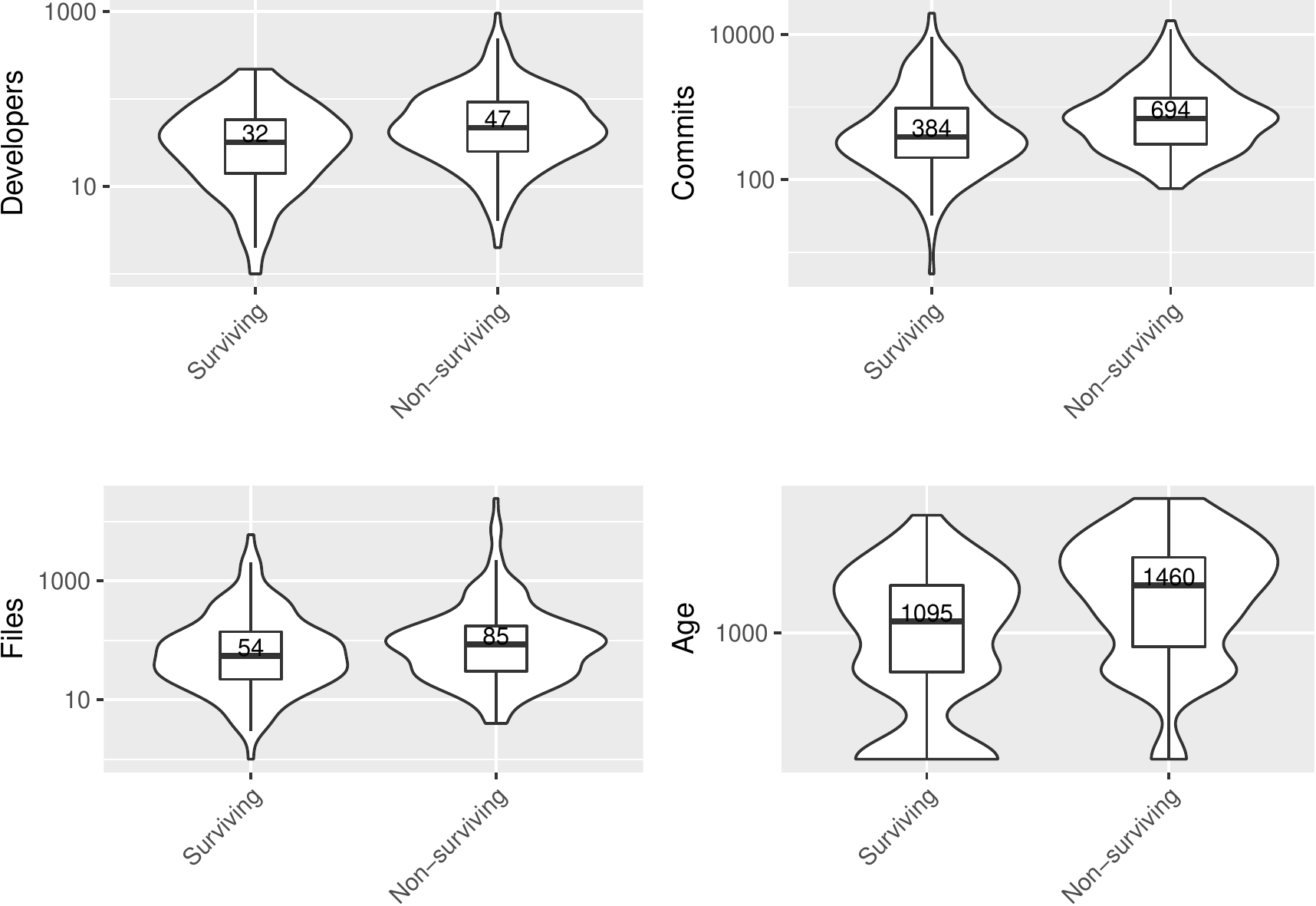}
    \caption{Number of developers, commits and files, and project age for surviving and non-surviving projects (at the date of the studied \tfdds)}
    \label{fig:versus}
\end{figure}

%After confirming the difference between the relative number of commits after the observed \tfdds of the surviving and non-surviving projects, we also compare them using other metrics. 
We also explore the differences, if any, between surviving and non-surviving systems w.r.t different factors in order to reveal if such factors can provide insights related to project survival. 
Figure~\ref{fig:versus} shows violin plots with the distributions of the number of developers, commits and files, and project age measured in days of the surviving and non-surviving projects. %at the date of the studied \tfdds. 
All values refer to the date of the studied \tfdds. 
We test the differences between surviving and non-surviving projects using two-sided Mann-Whitney tests and by visually confirming the differences using the visualizations of Figure~\ref{fig:versus}. 
Since we consider different aspects of the same projects we adjust the p-values to control for multiple comparisons using the method of Benjamini and Hochberg~\cite{Benjamini1995}. We select this method as it is more powerful than the alternative techniques.
%related methods over the false discovery rate. 
%source:https://towardsdatascience.com/an-overview-of-methods-to-address-the-multiple-comparison-problem-310427b3ba92

Interestingly, the surviving projects have less developers than the non-surviving ones (32 vs 47, median values, $p=2.2\times10^{-4}$). 
%p=0.0001112491
%ap =0.0002224982000
They also have less commits (384 vs 694, median values, $p=2.6\times10^{-4}$) and less files (54 vs 85, $p=4.7\times10^{-2}$).
%p=0.0001966424
%ap=0.0002621898667
%p=0.04790099
%ap=0.0479009900000
However, the effect size is \textit{negligible} for number of files ($d=0.13$) and \textit{small} for number of commits ($d=0.25$) and developers ($d=0.26$). 
Surviving projects are also younger at the time of the \tfdd (1095 vs 1460 days, median values, $p=3.4 \times10^{-7}$) with a \textit{medium} effect size ($d=0.37$). 
%p=0.00000008593478
%ap=0.0000003437391
We conjecture that non-surviving projects are either feature-complete (as they are more mature) or that they failed to attract new developers to assume their maintenance. However, it is important to consider that even feature-complete systems require corrective maintenance for fixing bugs~\cite{Talby2006}. It is thus uncommon that a project is both feature-complete and bug-free thus not requiring any further maintenance.

%\alex{discussion of Fig 10 might be biased as \% is determined relatively to the pre-TF-event history. What if the non-surviving systems are much older, so their pre-TF-event history is much longer and therefore the \% is much smaller even if the absolute number of commits is comparable? How is the comparison in Fig 10 affected by systems that might have survived one or more TF events but did not survive the last one?}

\begin{formal}
At the moment of the \tfdds, we found no major difference between surviving and non-surviving projects, in terms of number of developers, commits, and files. On the contrary, we found that surviving projects are younger at \tfdd time compared to the non-surviving ones.
%However, until such events, the difference between these projects \textit{small} (developers and commits) or even \textit{negligible} (files). 
\end{formal}

\section{Survey with new TF Developers}
\label{sec:survey}

%In the quantitative study described in the previous section, we did not find major differences in the surviving projects, when compared to the non-surviving ones using standard metrics available to GitHub repositories (commits, developers, and size). Therefore, to clarify this finding,

In this section, we report the results of a survey with the new TF developers of the surviving projects, i.e.,~developers that played a major role in the maintenance of these systems after the identified \tfdds, to answer to the last three research questions. 
The intention of \emph{RQ4} is to check whether the developers perceived the projects being at risk, before making the contributions that led them to reach a TF developer status.
\emph{RQ5} investigates how OSS managers should proceed to attract new TF developers to their systems. 
Finally, the intention of \emph{RQ6} is to shed light on programming and management practices that should be promoted (or avoided) in OSS development.

\subsection{Survey Design}
For each surviving project (128 projects), we select their {\em new} TF developers. 
After excluding  the ones without a public and valid e-mail address, we have identified 144 potential participants. 
We sent an e-mail to these developers with four questions: (1) Did you think that [project] was at risk of being discontinued before deciding to make major contributions to its continuation? (2) Why did you decide to make these contributions? (3) What project characteristics and practices helped you to make these contributions? (4) What were the main barriers you faced when making these contributions?

From the 144 e-mails we sent, four returned due to an invalid address. 
In total, we received 33 answers, representing a response rate of 24\%; %($33 / (144-4)$); 
this rate is higher than what has been achieved by previous papers~\cite{Palomba:TSE15,Vasilescu:CHASE}. 
To process the answers, we rely on Thematic Analysis~\cite{Cruzes2011}, which consists of (i) initial reading of the developer responses; (ii) generating initial codes for each response; (iii) searching for themes; (iv) reviewing the themes to find opportunities for merging among themes; and (v) defining and naming the final themes. 
%These steps were performed by the first and third authors of this paper; after that the final themes were checked and revised with the third author.
The three first steps were performed independently by two authors of this paper; after that, in a meeting, they performed the two last steps, by checking and revising the final themes.

\subsection*{RQ4) Do new TF developers perceive risks of project discontinuation?}
\label{sec:discontinuation}
We started the survey with the question about the perception of the discontinuation risks, according to the new TF developers. 
The purpose of this question is twofold. 
First of all, it serves as an additional validation of both our approach concerning the identification of \tfdds, i.e., whether or not \tfdds capture the project being at risk of discontinuation at the time, as well as the importance of the \tfdds: if developers believe that the project is at risk of discontinuation after a \tfdd, then further evolution of the project is indeed threatened. 
Second, this question assesses awareness of the new TF developers; awareness of the context and therefore of their tasks helps team coordination in software development~\cite{Espinosa2007TeamKA}.

Table~\ref{tab:rq1} summarizes the results for this question. 
Most respondents (18 developers, 60\%) agreed that the projects were facing risks of discontinuation. 
Examples of positive answers include: 
\answer{when I thought the project would die, I started making contributions to it once again}{D8}; and 
\answer{yes, otherwise the project would been completely abandoned}{D9}. 
Furthermore, we classify as a partial agreement five cases (17\%) where the developers reported problems in the projects, but were not clear about their severity, or mentioned the problem was mitigated by another developer who also entered in the TF set. 
As examples, we have these answers: 
\answer{[the] development had slowed}{D14}; and 
\answer{A new primary developer stepped in and took responsibility of the project after the original developer left.}{D32}. 
Indeed, the developer mentioned by D32 was also identified as a new TF developer in our study; therefore, we contacted him for our survey, but unfortunately did not receive a reply. 

Finally, six developers (20\%) answered that they did not perceive the projects as being at risk. 
Usually, these developers were succinct in their answers (just answering {\em no}, for example). 
Remarkably, among the negative answers, one developer mentioned the project is supported by a major software company, which contributes to reduce the discontinuation risks, in his  opinion: 
\answer{this open source project is actually backed by a for profit company, so the project didn't risk being abandoned}{D24}. 
Finally, we classify one answer as unclear (3\%), because it is not related to the provided question. 
Four respondents did not answer this survey question. 

\begin{table}[!t]
 \caption{Did you perceive the projects at risk?}
 \centering \small
 \scalebox{0.9}{
 \begin{tabular}{ c  c c c}
 \toprule
 {\bf Yes} & {\bf Partially} & {\bf No} & {\bf Unclear}\\ 
 \midrule
 18 (60\%) & 5 (17\%)  & 6 (20\%) & 1 (3\%) \\  \bottomrule
 \end{tabular}
 }
 \label{tab:rq1}
\end{table}

\begin{formal}
77\% of the new TF developers were (partially) aware of the risks faced by the surviving systems, before making the contributions responsible for the project recovery.
\end{formal}

\subsection*{RQ5) What motivates a developer to assume an open source project after a \tfdd situation?}
\label{sec:motivations}

\begin{table}[!t]
  \centering
  \small
  \caption{Motivations to contribute}
  \label{tab:answers_q2}
   \scalebox{0.85}{
  \begin{tabular}{lrr}
  \toprule
  \multicolumn{1}{l}{\bf Motivations} & {\bf Devs}  & {\bf \%} \\
  \midrule
  Because I was using the project              & 17 &   53 \\ 
  To contribute to an open source project      & 11 &   34 \\
  To avoid the project discontinuation         & 5 &   	16\\
  I have interest on the project area          & 4 & 	13\\
  I get paid to contribute              	   & 4 & 	13\\
  To improve my own skills                 	   & 3 &  	9\\ 
  I have the skills required by this project  & 3  &     9\\ 
  It is a successful project                  & 3  & 	9\\
  Others                					  & 7 &  	22\\
  \bottomrule
  \end{tabular}
  }
\end{table}

%\noindent\textit{Question \#2. Why did you decide to make these contributions?}

With this question, we aim to reveal the reasons that motivated the new TF developers to make their major contributions to the projects. 
Table~\ref{tab:answers_q2} summarizes the main reasons mentioned by the surveyed developers. 
\textbf{Because I was using the project to address my personal or professional needs} is the most common reason mentioned by 17 participants (53\%). 
Answers referring to this reason include
\answer{I used the [project] in my own products and was struggling with a few bugs so I decided to fix them and contribute back}{D14}; 
\answer{mostly because I used [project] heavily and was asked for documentation and improvements from within my company}{D16}; and 
\answer{I used the project professionally and was in a position to provide some level of support as part of my job}{D23}. 
\textbf{To contribute to an open source project} is the second most common reason mentioned by 11 participants (34\%); 
\textbf{to avoid the project discontinuation} is mentioned by five developers (16\%), e.g.,
\answer{I was the only additional contributor on the project so I was the only person capable of keeping it alive}{D7}; 
%and
%\answer{it will die if I don't [contribute], and I think it still has value.}{D13}. 
% The remaining reasons mentioned by the participants are as follows: 
% \textit{I have interest on the project area} (4 developers),
% \textit{I get paid to contribute} (4 developers), 
% \textit{to improve my own skills} (3 developers), 
% \textit{I have the right skills to contribute to this project} (3 developers), 
% \textit{it is a successful project} (3 answers), 
% \textit{it is a promising or interesting project} (2 answers),
% \textit{I was invited to contribute} (2 answers),
% \textit{for personal satisfaction} (2 answers), and
% \textit{to attract developers to my company} (1 answer).
One respondent did not answer this question. 

\begin{formal}
The developers responsible to reactivate the maintenance of the surviving projects were motivated by their own usage of the projects (17 developers, 53\%). They also intended to contribute back to an open source community (34\%) or avoid the project discontinuation (16\%).
\end{formal}
% %%%%%%%%%%%%%%%%%%%%%%%%%%%%%%%%%%%%%%%%%%%%%%

\subsection*{RQ6) What project characteristics most facilitate or hamper the work of recently arrived TF developers?}
\label{sec:enablers-barriers}

\begin{table}[!t]
\centering
\small
\caption{Characteristics that helped new TF developers}
\label{tab:question3_answers}
\scalebox{0.85}{
\begin{tabular}{llrr}
\toprule
\multicolumn{1}{l}{\textbf{Type}}                          & \textbf{Characteristics}                                        & \textbf{Devs} & \textbf{\%} \\
\midrule
%\multicolumn{1}{l}{\multirow{2}{*}{Human}}
Human/ & Friendly and active owners/members                & 12  & 41
\\
Social & I liked/knew the project           & 3  & 10	\\
\midrule
\multirow{5}{*}{Technical}                                
                                                           & Programming language                                   & 4  & 14         \\
                                                           & Well-known SE principles  	   & 4  & 14        \\
                                                           & Pull based development  	   & 4  & 14        \\
                                                           & Continuous integration                                  & 2  & 7         \\
                                                           & Clean and well-designed code                    & 2  & 7        \\
                                       %                    & Easiness to contribute                                 & 1             \\
                                       
%                                  \midrule
%\multirow{4}{*}{Process}                                   & Pull-based development                                    & 4             \\
                                                           & Code revision              & 1    & 3      \\
         
\midrule
\multirow{4}{*}{Others}   & Main repository access                                   & 3 & 10 \\
                    &  Job support                                & 2 & 7           \\  

     & Small or simple project                                           & 2 & 7          \\  
                    & Open source license                                            & 1  & 3 \\           											   
\bottomrule                                                           
\end{tabular}
}
\end{table}

We start with the project characteristics that facilitated the attraction of the new TF developers. As presented in Table~\ref{tab:question3_answers}, we organize these characteristics in three groups: human and social characteristics (15 answers), technical characteristics (17 answers), and other characteristics (8 answers).
The most mentioned human and social characteristic is the presence of friendly and active project owners or members (12 answers). 
As examples, we have these answers: 
\answer{it has been [dev-name]'s kindness to my first contributions and his help to me, and later other cool developers' support}{D6}; 
\answer{the responsiveness of the existing maintainer was the key factor to my ongoing contributions}{D11}. 
Among others, technical characteristics include the usage of a specific programming language (4 answers) or following well-known software engineering principles and practices (4 answers). 
The last category groups factors like permission to access the main repository (3 answers) and financial support by  a company (2 answers). 
Four respondents did not answer this question.

\begin{formal}
The characteristics that helped on the attraction of new TF developers have a social, technical or external nature. Friendly and active maintainers is the most mentioned facilitator, indicated by 12 developers (41\%).
\end{formal}

\begin{table}[!t]
\centering
\small
\caption{Barriers faced by new TF developers}
\label{tab:question4_answers}
\scalebox{0.85}{
\begin{tabular}{llrr}
\toprule
\multicolumn{1}{l}{\textbf{Type}}                          & \textbf{Barriers}                                        & \textbf{Devs} & \textbf{\%} \\
\midrule
\multicolumn{1}{l}{\multirow{3}{*}{\shortstack[l]{Human/\\Social}}} 
& Lack of time                & 7  & 26
\\
& Lack of experience          & 3   & 11   \\
& Unfriendly maintainers      & 2   & 7	\\
\midrule
\multirow{2}{*}{Technical}                                
&  Need to keep backward compatibility      & 4   & 15     \\                                                                                   & Lack of well-known SE principles          & 1   & 4      \\
\midrule
\multirow{3}{*}{Others}   
& Lack of access to the main repository     & 5 & 19 \\
& Large number of pending issues  			& 3 & 11 \\
&  No financial support                     & 2 & 7  \\  				
\midrule
No barriers & \multicolumn{1}{c}{-} 	& 4 & 15\\
\bottomrule                                                           
\end{tabular}
}
\end{table}

To complement the answer to \textbf{RQ6}, we also asked the new TF developers about the barriers they faced when making the contributions that led them to achieve a status of TF developer. 
As in the case of the first part of the question, we organize the answers mentioned by the participants in three groups: human and social barriers (12 answers), technical barriers (5 answers), and other barriers (10 answers). 
Table~\ref{tab:question4_answers} presents the answers in each group. 
As we can observe in this table, most answers denote human and social barriers.
Particularly, {\em lack of time} is the most common barrier mentioned by the survey participants (7 answers). 
As examples, we have these answers: 
\answer{I have other projects to maintain.}{D3}; and 
\answer{time is always an issue, especially because the range of features is fairly wide}{D23}. 
Technical barriers include the requirement to {\em keep backward compatibility and do not introduce bugs} (4 answers) and the lack of solid software engineering principles (1 answer). 
Another barrier commonly mentioned by the participants is the {\em difficulty to obtain access to the main repository} (5 answers). 
The participants justify the need to obtain this access because the {\em maintainers are absent} (D2, D12) or {\em the project was abandoned} (D8, D9). 
Four developers mentioned they faced no barriers at all. Six developers did not answer this question.
\begin{formal}
Human and social barriers are the most common ones faced by new TF developers; particularly, lack of time is the most common barrier.
\end{formal}

%\input{ecosystems}
%!TEX root = paper.tex
\section{Discussion}
\label{sec:discussion}
%In this section, we summarize and discuss the relevance of our study findings.\\[-0.3cm]

\noindent{\em Truck factor is not only a theoretical metaphor:} In OSS development, it is possible to argue that TF is just a theoretical scenario, since the code is public and others can assume the maintenance work if the key developers abandon the project. In fact, one of the participants of the survey provides an argumentation in this direction: \answer{it's open source, if people want to use it, they will use it. If it's missing features they really want/need, they will submit PR's, or fork and maintain their own copy.} {D30}. Undoubtedly, if the code is public on GitHub, anyone has the legal permission (according to the project's license) to collaborate with or fork the project. 
Moreover, GitHub provides many useful instruments to facilitate this process, like easy forking or pull requests. Despite that, our study shows that even popular projects may fail to attract new contributors after being abandoned. More precisely, only 41\% of the projects  have fully recovered the maintenance activity after the \tfdds studied in our work. We hypothesize that assuming the maintenance of an open source project is a complex task, which requires time, technical and social skills and familiarity with the project domain. Many projects therefore do not succeed to find developers with this profile and face serious maintenance problems or even fail after being abandoned by their TF developers. 
We stress that projects cannot rely only on peripheral contributions to survive as such contributions are complementary to the ones of TF developers and cannot fully support the project maintenance \cite{Setia2012,Terceiro2010}.

Interestingly, we also found arguments in the opposite direction, stating that TF is a less important concern in software projects backed by a company regardless of the project being open source. We received at least two answers hinting in this direction, as this one: \answer{Most of the questions are not relevant because [project] is actually a large project with formal sponsorship
by [company]}{D33}. In other words, these developers consider that TF is a real concern only in projects without financial support, as is the case of most open source projects.

Regardless of the point of view, our work showed the implications of TF developer abandonment in project evolution. It is therefore essential, for project key maintainers either to strive to increase the number of TF developers or to provide an alternative backing, e.g., company-based support, in order to prevent or reduce the chances of \tfdds. 
%Based on the replies of the survey, strategies to achieve this include engaging with existing peripheral developers within the project or newcomers, given their willingness to take on more responsibilities.
\\[-0.4cm]

\noindent{\em How to overcome \tfdds:} Although we show that \tfdd situations are a reality in OSS development, we also found that it is  possible to survive and recover the maintenance after attracting new developers to the TF set. By surveying these new TF developers, we shed light on two key characteristics of the surviving projects. 

\textit{First}, the surveyed developers decided to assume the maintenance of these projects motivated by their own needs, since they were using the projects and require new features or fix existing bugs. This finding suggests a connection between the number of users of an OSS project and its resilience to \tfdds. Particularly, 53\% of the TF developers surveyed in our study were attracted because they were earlier users of the projects and therefore had personal interests in avoiding their failure. 

\textit{Second}, human and social factors have a key role on attracting new TF developers. According to the survey participants, 51\% of the factors that helped in their attraction are social in nature; and 44\% of the barriers faced in this process are also human and social ones. The importance of human and social barriers to technical contributions was also observed by Palomba et al.~\cite{Palomba:ICSE18}. 
Our findings, therefore, confirm the importance of human and social factors in OSS development. This is particularly the case if most contributions are voluntary, as indicated by one of the survey respondents: \answer{There is no authority over the top that chooses who will work on what. We are all contributing during our ``free'' time, for only the ``enjoyment'' of it. So, we sort of contribute only where it ``feels'' good}{D26}.

Our work sheds light on the origin of new TF developers that assume project maintenance. 
In turn, this information can be utilized to prevent or limit the prevalence of \tfdds. % by identifying individuals to join the TF developers of a project. 
TF developers that consider leaving a project, can identify individuals that can serve as potential new TF developers and prevent the discontinuation of the project. 
Our study shows that such individuals can be either volunteers using the project or company-based developers that have had prior contributions to the project, in the case that a company uses the project.

%This motivation was mentioned by xx\% of the survey participants. 
%Interestingly, we also found arguments in the other direction: if the (open source) project is supported by a company, it will survive due to its access to financial support. As example, we have this answer: 
%!TEX root = paper.tex
\section{Threats to validity}
\label{sec:threats}
\subsubsection*{Construct validity}
%\noindent\textit{Construct Validity:}
Firstly, our results depend on the accuracy of TF computations. 
To mitigate this threat we used the TF algorithm that presents the best accuracy, as pointed by a recent comparative study~\cite{Ferreira2017}.
Nonetheless, we quantified the risk stemming from the choice of the TF algorithm by measuring the distribution of the percentage of files of the new TF developers for the 128 surviving systems. 
If the new TF developers work on few files, then they should not be considered TF developers suggesting imprecision of the algorithm.
We observed that 
%the 1st, 2nd and 3rd quartiles equal to 27\%, 41\% and 60\% of the systems' files with minimum and maximum values of 9\% and 100\% respectively. Thus, 
new TF developers have worked on a substantial share of a system's files ($Q_2 = 41\%$) increasing our confidence in the algorithm.

Identification of \tfdds depends on the selected abandonment threshold. 
%\deleted{However, there is no consensus in the literature on appropriate thresholds to identify such developers, e.g., Constantinou and Mens~\cite{Constantinou:2017:ISSE} used a 1-year threshold as well, while Lin et al.~\cite{Lin2017} used a 180-day threshold to find developers abandoning a project. Since this study focuses on key developers, we believe that a less strict threshold can reduce the effect of this threat in our analyses.}
As there is no consensus on this threshold in the literature, we experimentally test five different thresholds (Section~\ref{sec:threshold}) to select an appropriate threshold. 
However, using a different threshold might lead to different results. 
The TF measures may also vary due to possible developer aliases. We mitigate such a threat by carefully handling common sources of aliasing (Section~\ref{sec:aliases}).

%\added{
Another threat concerns our definition of \tfdd as occurring if \emph{all} TF developers abandon the project. 
%However, any type of automatic validation of our definition will inevitably rely on heuristics, and can also be prone to errors. 
To evaluate the accuracy of this definition, we rely on the survey. %responses from TF developers who assumed the project maintenance. 
While we are aware of the limitations of this evaluation strategy, we observe that more than 75\% of the respondents confirmed their (partial) awareness that the project was indeed at risk.
Therefore, we conclude that the effect of this threat is low.
%}

Finally, we see a system as an evolving artifact, requiring continuous change. For mature software that does not require changes, a \tfdd analysis can thus lead to erroneous results. 
However, this is uncommon as based on Lehman's software evolution laws~\cite{Lehman1980}, software must be continuously adapted or it becomes progressively less satisfactory. 
To quantify this threat we manually inspected the README files of the 187 non-surviving systems. 
Although we found one system (\repo{defunkt/jquery-pjax}) described as feature-complete, it is still maintained when bugs need to be fixed.
%``it might continue to receive important bug fixes, but its feature set is frozen'' indicating the need for future maintenance.
% \eleni{I suggest to add here:\\
% Nonetheless, we quantified this threat by manually inspecting the READMEs of the 187 non-surviving systems and although we only found 1 system\footnote{\url{https://github.com//defunkt/jquery-pjax}} that explicitly states that it is feature-completed, it is also mentioned that "it might continue to receive important bug fixes, but its feature set is frozen" indicating the need for future maintenance.
% }
% \guilherme{I agree}
%}

\subsubsection*{Internal Validity}
Identification of \tfdds uses data from the entire development history of a project, and is, hence, sensitive to the partial loss of history, e.g., due to corrupted migration from another version control system. 
We mitigate this threat by excluding projects with evidence of a corrupted migration to GitHub  (Section~\ref{sec:dataset}). We also exclude non-software projects, such as books and tutorials.
%\eleni{what do we mean by corrupted migration?}
%\guilherme{I gave an example to illustrate}

%We consider that all files have the same importance when computing the truck factor. We cannot know for sure that a file is needed to be maintained or not. Therefore, if a project has many files that do not require changes TF leavers who had worked on those files will still appear as TF developers either way. This situation can mask a project recover.

\subsubsection*{External Validity} Our dataset was carefully selected from popular projects on GitHub.
%, coming from six different programming languages. 
However, our findings cannot be generalized to other projects and particularly to closed-source projects. Indeed, our survey results suggest that \tfdds in the context of software with financial support might have very different characteristics. %\\[-0.3cm]
%Despite these observations, our findings---as usual in empirical software engineering---cannot be directly generalized to other projects, mainly closed-source ones.

%!TEX root = paper.tex
\section{Related work}
\label{sec:rw}
%\alex{Cite \emph{Nassif and Robillard, "Revisiting Turnover-Induced Knowledge Loss in Software Projects", ICSME 2017} as recommended by the ICSME reviewers.}
%\guilherme{I added this new reference in the second paragraph.}
Truck factor is a concept defined by the agile community to assess knowledge concentration in software projects. As the concept initially lacked a formal definition, the first works in this area focused on proposing algorithms to compute truck factors.
The first algorithm to this purpose was proposed by Zazworka et al.~\cite{Zazworka2010}. After that,  it was used by Ricca et al.~\cite{Ricca2010} and  Torchiano et al.~\cite{Torchiano2011}, respectively, to investigate the presence of ``heroes'' in open source projects and to investigate threshold values to use when computing truck factors. However, Zazworka's algorithm suffers from scalability problems~\cite{Ricca2011,Hannebauer2014}, which limits its applicability to real systems. 
To address these problems, new algorithms were proposed by
Cosentino et al.~\cite{Cosentino2015},
%proposed a hierarchical algorithm, which aggregates file-level authorship results to modules and, in a second step, aggregates module-level results into systems; 
Rigby et al.~\cite{Rigby2016} 
%proposed a solution inspired by a Monte Carlo simulation algorithm; 
and Avelino et al.~\cite{Avelino2016}.
%use code authorship metrics to identify source code files' authors~\cite{Fritz2010, Fritz2014}. Essentially, Avelino's algorithm relies on a greedy approach to identify the developers that together control the authorship of most files in a system.
Ferreira et al.~\cite{Ferreira2017} compared these three algorithms and concluded that the latter algorithm is the most accurate one. 
None of the aforementioned works investigated whether \tfdds really occur and what happens with open source projects afterwards. 

Truck Factor can be considered as a particular case of turnover, involving the principal developers of a project. Turnover of developers in general is a well-studied phenomenon in software 
engineering.  Foucault et al.~\cite{Foucault2015} report the negative impacts
of turnover in the internal quality of five open source projects. Rigby et al.~\cite{Rigby2016} and later Nassif et al.~\cite{Nassif2017}, profiled the knowledge loss induced by developers who leave a software project and provide tools to help large projects to assess the risk of turnover.
Hilton and Begel~\cite{Hilton2018} recently studied internal turnover in 
a major software company, with more than 30K employees. 
By surveying a sample of 374 of such employees, they reveal what causes
engineers to consider leaving their teams, why they leave, how
they learn about new teams, and how they decide which team to
join. Lin et al.~\cite{Lin2017} conducted a similar study, but with focus on 
five open source projects. They show that developers are retained when they (i) start contributing to 
the projects earlier, (ii) maintain both code developed by others 
and their own code, and (iii) mainly code instead of writing documentation. 
Vasilescu et al. have shown that presence of developers with very different durations of engagement on GitHub increases the likelihood of turnover~\cite{VasilescuPRBSDF15}.
Qiu et al. have observed that while women tend to disengage from open source projects faster than men, attachment of women to open teams with regard to diversity of information increases their chance of prolonged engagement more than the chances of men~\cite{QiuNBSV19}. 
Recently studies of contributor disengagement have been conducted by Miller et al.~\cite{MillerWKV19} and  Iaffaldano et al.~\cite{abs-1903-09528}.

Motivations and barriers to contribute to open source systems were previously investigated for different developer profiles:
developers that have only one patch accepted~\cite{Lee2017}, developers with few contributions and no intention to become an active project member~\cite{Pinto2016}, newcomers~\cite{Steinmacher2015, Steinmacher2016}, and core developers~\cite{Coelho2018}. 
Regarding the reasons that motivate developers to contribute to open source, some of these studies also show that core developers are motivated by their personal needs, as we concluded for the specific case of new TF developers. By contrast, one-time contributors and casual contributors are mainly motivated by the need to fix minor bugs.
Lack of time is a common barrier to contribute, mentioned by core developers, one-time contributors, and casual contributors. It was also commented by the new TF developers surveyed in our study. 
Steinmacher et al.~\cite{Steinmacher2015} defined a conceptual model composed of 58 barriers that may hamper newcomers' first contributions. 
They list and classify these barriers, but do not provide insights on their frequency. 
Coelho et al.~\cite{Coelho2017}, report a survey with 
the maintainers of 104 failed open source projects, i.e.,~projects that are not maintained anymore. According to their survey, the most common reasons for open source project failures are the appearance of a strong competitor, obsolescence, and lack of time or interest of the project owners.
By comparing the factors that attract new contributors with our results about attracting new TF developers, we found that new TF developers are motivated by their own use and need to save the projects (53\%) in contrast to new, casual or one-time contributors. On the other hand, our results about the importance of human and social barriers in OSS comply with related work, e.g., reception issues~\cite{Steinmacher2015} and friendly and active owners/maintainers (12 out of 29 respondents in our survey).
Identification of welcoming projects with friendly maintainers is related to such topics as community health~\cite{2018soheal}, presence of codes of conduct~\cite{TouraniAS17} and emotions expressed in developer communication~\cite{GachechiladzeLN17}.
%Finally, the importance of human and social factors to facilitate the contributions is also mentioned in the core developers and one-time contributors studies.

%\alex{Extra literature to check: the fact that TF events happen early might be related to some management/organizational science findings about loss of interest in projects, group dynamics etc. Barriers: there is work of Denae Ford on barriers to participation in StackOverflow + recent ICSE SEIS paper of David Redmiles on barriers in GitHub---however, they focus on gender differences. Still there maybe something related. }
%!TEX root = paper.tex
\section{Conclusion}
\label{sec:conc}

In this paper, we presented an in-depth investigation of the occurrence of \tfdds in open source projects, i.e., the abandonment of a project by its principal developers. 
We showed that \tfdds are not only a metaphor, but they indeed happen in open source projects (in 16\% of such projects, at least in our sample of 1,932 GitHub projects). 
Additionally, we showed that projects survive such situations, by attracting new core contributors (41\% of the projects survived a \tfdd, in our sample). 
Finally, we reveal the motivations that led these developers to take over the studied projects, after the projects faced a \tfdd. We also reveal the principal enablers and barriers faced by these developers during this process. 
%This list of enablers and barriers can be used by project leaders to improve the management practices employed in their projects. 
This list of enablers and barriers are especially useful and should be considered to build development communities that are more attractive to new contributors.

As future work, we envision the design, implementation and evaluation of tools to assess the risks faced by an open source project, in case it is abandoned by its TF developers. This assessment is particularly important to the users of such projects. We also see space to investigate recommenders of TF developers for a system, based for example on their own usage of the projects. 
%Finally, we think that the successful cases of  overcoming TF events, as reported in this paper, should be advertised in open source communities, to show to their members that assuming the maintenance of an open source project is not an unthinkable possibility. 
Finally, open source communities should be made aware of successful cases of projects overcoming \tfdds, as we report in our paper, and motivate developers to actively contribute to projects at risk. 

As a final note, we provide a replication package with the results of our analysis as well as the survey's answers, repositories data, and scripts used in the paper. The replication package is available at \url{https://doi.org/10.5281/zenodo.2546008}.

\section*{Acknowledgments}
\noindent We thank all respondents of our survey. This work was partially supported by the Excellence of Science Project SECO-Assist (O015718F, FWO - Vlaanderen and F.R.S.-FNRS), the FRQ-FNRS collaborative research project R.60.04.18.F SECO-Health, CAPES (131987/2016-01) and CNPQ (140205/2017-9).

\newpage

\balance
\bibliographystyle{IEEEtran}
\bibliography{references}

\end{document}